\documentclass[12pt,a4paper]{article}
\usepackage{latexsym}
\usepackage{epsf}
\def\unit{\hbox to 3.3pt{\hskip1.3pt \vrule height 7pt width .4pt \hskip.7pt
\vrule height 7.85pt width .4pt \kern-2.4pt
\hrulefill \kern-3pt
\raise 4pt\hbox{\char'40}}}

\def\x{\times}

\begin{document}

\begin{flushright}
\footnotesize
UG-3/99\\
SPIN-99/03 \\
February, $1999$
\normalsize
\end{flushright}

\begin{center}


\vspace{.6cm}
{\LARGE {\bf Kaluza-Klein Monopole and 5-brane Effective Actions}}
\footnote{To appear in the proceedings of the {\sl 32nd Symposium
Ahrenshoop on the Theory of Elementary Particles}, Buckow, Germany,
September 1998.}

\vspace{.9cm}

{
{\bf Eduardo Eyras}
\footnote{E-mail address: {\tt E.A.Eyras@phys.rug.nl}}\\
{\it Institute for Theoretical Physics\\
University of Groningen \\
Nijenborgh 4, 9747 AG Groningen, The Netherlands}
}

\vspace{.2cm}

{and}

\vspace{.2cm}

{
{\bf Yolanda Lozano}
\footnote{E-mail address: {\tt Y.Lozano@phys.uu.nl}}\\
{\it Spinoza Institute\\
University of Utrecht\\
Leuvenlaan 4, 3508 TD Utrecht, The Netherlands}
}

\vspace{.4cm}

\vspace{.8cm}


{\bf Abstract}

\end{center}

\begin{quotation}

\small

We review the construction of the Kaluza-Klein
monopole of the Type IIA theory in the most general case of
a massive background, as well as its relation via T-duality
with the Type IIB NS-5-brane. This last effective action is
shown to be related by S-duality to the D5-brane effective action.

\end{quotation}

\vspace{1cm}

\newpage

\pagestyle{plain}

\newpage
                                        
\vskip 12pt
\begin{figure}[!ht]
\begin{center}
\leavevmode
\epsfxsize= 10cm
\epsffile{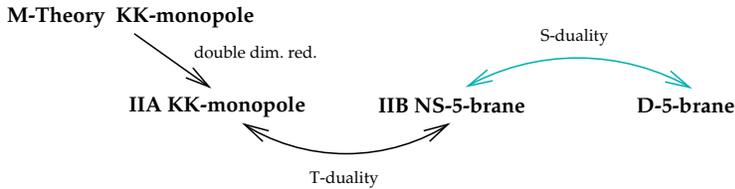}
\caption{\footnotesize
Derivation of the different actions 
considered in this paper. The (massive) IIA KK-monopole is obtained
from the (massive) M-theory KK-monopole after a double dimensional reduction.
Applying now $T$-duality in the IIA KK-monopole action 
the Type IIB NS-$5$-brane
action is obtained, and it is $S$-dual to the D5-brane effective action.}
\label{fig:cuadro2}
\end{center}
\end{figure}

The most general Type IIA superstring theory contains a mass 
parameter dual to
the RR 9-form potential to which the D8-brane couples minimally
\cite{BRGPT}, the corresponding low energy effective action being the
massive IIA supergravity of Romans \cite{Romans}. An eleven
dimensional interpretation of this theory can only be found when
the eleventh direction is compact \cite{BLO}, 
with Lorentz invariance taking place
in the other ten dimensional coordinates\footnote{This circumvents the no-go
theorem of \cite{BDHS}.}. The explicit eleven dimensional supergravity
action depends on the Killing vector associated to translations along
the eleventh coordinate, and gives the massive Type IIA supergravity
action of Romans after a direct dimensional reduction
along this direction.
The effective actions of the M-branes of this theory are described by
gauged sigma-models in which the Killing isometry is gauged, with a 
gauge coupling constant proportional to $m$.
Two massive branes in the Type IIA theory can be obtained from any such
M-brane depending on whether we perform a direct or a double dimensional
reduction. In particular the massive D6-brane is obtained by
dimensional reduction along the Taub-NUT direction of a massive 
M-KK-monopole in which the two Killing vectors associated to the mass
and the Taub-NUT isometries coincide \cite{BEL}. 
The construction of a massive KK-monopole in eleven dimensions
is more subtle than that of an ordinary brane, since the monopole
is already described by a gauged sigma-model in the massless case.
The massless M-theory KK-monopole \cite{BJO} behaves like a 6-brane,
and its field content is that of the 7-dimensional vector multiplet,
involving 3 scalars and 1 vector. Since the embedding coordinates
describe 11-7=4 degrees of freedom one scalar has to be eliminated
by gauging an isometry of the background. This isometry is that
associated to translations of the Taub-NUT direction of the 
transverse space to the monopole.
The massive M-KK-monopole of \cite{BEL} cannot give
rise to a massive Type IIA KK-monopole after double dimensional
reduction, since the gauged isometry disappears in the reduction.
One needs to construct a massive M-KK-monopole in which the two isometry
directions associated to the mass and the Taub-NUT space are different.
Double dimensionally reducing along the direction associated to the
mass will give rise to the effective action of the massive IIA
KK-monopole. This explicit effective action has been constructed in \cite{EL}.
Here we just summarize its worldvolume field content and the 
interpretation of these fields in terms of solitons in the brane.

\begin{table}[h]
\renewcommand{\arraystretch}{1.5}
\begin{center}
\begin{tabular}{|c||c|c|c|c|c|c|}
\hline
M-theory KK  & $ $ & ${\hat \omega}^{(1)}_{{\hat \imath}}$ &
${\hat d}^{(5)}_{{\hat \imath}_1 \dots {\hat \imath}_5}$ & $ $ &
${\hat \omega}^{(6)}_{{\hat \imath}_1 \dots {\hat \imath}_6}$
& ${\hat \omega}^{(7)}_{{\hat \imath}_1 \dots {\hat \imath}_7}$  \\
\hline
IIA KK & $\omega^{(0)}$ & $\omega^{(1)}_i$ & $d^{(5)}_{i_1\dots i_5} $ &
$\omega^{(5)}_{i_1\dots i_5}$ & $\omega^{(6)}_{i_1\dots i_6}$ 
& $ $ \\
\hline
\end{tabular}
\end{center}
\caption{\label{table-mMKK} \small {\bf Worldvolume field content of the
massive M-theory and Type IIA KK-monopoles.} In this table we summarize the
worldvolume fields present in the effective actions 
of the massive M--theory and massive IIA KK--monopoles.}
\renewcommand{\arraystretch}{1}
\end{table}

We find, as in the massless case, a 1-form ${\hat \omega}^{(1)}$
describing a 0-brane soliton in the worldvolume of the KK-monopole.
Its dual 4-form describes a 3-brane soliton. 
They correspond to the intersections: $(0|{\rm M}2,{\rm MKK})$ 
and $(3|{\rm M}5,{\rm MKK})$,
respectively. 
The monopole contains as well a 
4-brane soliton which couples to the 5-form dual to one of the embedding
scalars:\\ $(4|{\rm MKK},{\rm MKK})_{1,2}$. 
All these intersections have been already discussed in the literature for the
massless monopole \cite{Papa}.

In the massive case we find a new field ${\hat d}^{(5)}$ which 
couples to the 5-brane
soliton represented by the configuration\footnote{We use here a 
notation where $\times (-)$ indicates
  a worldvolume (transverse) direction. The first $\times$ in a
  row indicates the time direction. The $z$-direction in the monopole 
corresponds to the
isometry direction of the Taub-NUT space. A single M9-brane
contains as well a Killing isometry in its worldvolume, as has been
discussed in \cite{BvdS,proci}. This 
isometric direction has been depicted as well as a $z$-direction.}:

\begin{displaymath}
 (5|{\rm M}9,{\rm MKK})=
\mbox{ 
$\left\{ \begin{array}{c|cccccccccc}
         \x & \x & \x  & \x & \x & z & - & \x & \x  & \x  & \x \\
         \x & \x & \x   & \x  & \x & \x & \x & z & -  & - & -           
                                    \end{array} \right.$}   
\end{displaymath}

\noindent The 5-brane soliton predicted by the M-KK and M9-brane worldvolume
supersymmetry algebras is realized as a 4-brane
soliton given that it cannot develop a worldvolume direction along the
isometry of the M9-brane. This is in agreement with the worldvolume
field content that we have found for the massive M-KK-monopole, since
the only worldvolume field to which this soliton can couple
is the 5-form ${\hat d}^{(5)}$.

The 6-form ${\hat \omega}^{(6)}$ is interpreted as the tension of
the monopole and couples to the 5-brane soliton
realized as the embedding of an M5-brane on the
KK-monopole.
Finally, ${\hat \omega}^{(7)}$ couples to the 6-brane soliton describing
the embedding of the monopole in an M9-brane: $(6|{\rm M}9,{\rm MKK})$.
{}From the point of view
of the M9-brane this 6-brane soliton couples to the dual of its 1-form
vector field \cite{proci}.        

The double dimensional reduction of the action of the massive M-theory
KK-monopole gives that of the massive IIA KK-monopole, and it is constructed
explicitly in \cite{EL}. Its worldvolume field content is summarized
in Table 1.

The worldvolume fields that couple
to the soliton solutions of a KK-monopole are those necessary to
construct invariant field strengths for the fields
$i_k C^{(p+1)}$ (see \cite{EJL}). 
These field strengths have the form:

\begin{equation}
{\cal K}^{(p)} = p\partial\omega^{(p-1)}+\frac{1}{2\pi\alpha^\prime}
(i_k C^{(p+1)})+\dots  \, ,
\end{equation}

\noindent so that $\omega^{(p-1)}$ couples to a
$(p-2)$-brane soliton which describes the boundary of a $p$-brane
ending on the monopole, with one of its worldvolume directions
wrapped around the Taub-NUT direction of the monopole. The
target space field associated to $\omega^{(1)}$ is $(i_k C^{(3)})$
so that it describes a wrapped D2-brane ending on the monopole.
It is worth noting that the Born-Infeld field does not couple to the
action, which reflects the fact that there are no fundamental strings
ending on the monopole.

As in the massless case, we find the configurations: 
$(2|{\rm D}4,{\rm KK})$, $(3|{\rm NS}5,{\rm KK})$ and 
$(3|{\rm KK},{\rm KK})_{1,2}$ \cite{Papa}. 
When the mass is non-zero there is a new worldvolume field 
$d^{(5)}$ (associated to dual massive
transformations), which couples to a 4-brane soliton.
This soliton is a domain wall in the six dimensional
worldvolume and is described by the configuration:

\begin{displaymath}
 (4|{\rm D}8,{\rm KK})=
\mbox{ 
$\left\{ \begin{array}{c|ccccccccc}
         \x & \x & \x  & \x & \x & - & \x & \x & \x  & \x \\
         \x & \x & \x   & \x  & \x & \x & z & - & -  & -           
                                    \end{array} \right.$}   
\end{displaymath}

\noindent which is obtained by reducing the $(4|{\rm M}9,{\rm KK})$ soliton
configuration of the M-theory KK-monopole along the isometric
direction of the M9-brane.
This intersection is related by T-duality to various intersections
considered in the literature.
There is as well another 4-brane soliton in the KK-monopole
worldvolume which is already present in the
massless case. This is the embedding of the D4-brane on
the monopole: $(4|{\rm D}4,{\rm KK})$, and it couples to the worldvolume
field $\omega^{(5)}$, describing the tension of the monopole.
Finally, the reduction of the 6-brane soliton $(6|{\rm M}9,{\rm KK})$ 
gives a 5-brane soliton coupled to $\omega^{(6)}$ which is
realized as the embedding of the KK-monopole on a
KK-7A-brane\footnote{This brane is obtained by reducing the M9-brane
along a worldvolume direction other than the $z$-direction,
but it is not predicted by the Type IIA supersymmetry algebra
\cite{Hull}.
As discussed in \cite{proci}, this is also the case for the 
KK-6A brane, obtained by reducing the M-KK-monopole along a
transverse direction different from the Taub-NUT direction.
These branes are required by U-duality of M-theory on a 
d-torus (see for instance \cite{OP}).}.

We can now construct the effective action of the Type IIB 
NS-5-brane by performing a (massive) T-duality transformation
in the IIA KK-monopole along its Taub-NUT direction. The details
of this calculation are presented in \cite{EJL}. 
The result is the following effective action:
\begin{equation}
\label{5brane-action}
\begin{array}{rcl}
S &=&
-T_{{\rm B5}} \int d^6 \xi \,\,
e^{-2\varphi} \sqrt{1 + e^{2 \varphi} (C^{(0)})^2}
\sqrt{|{\rm det} 
\left( g - (2\pi\alpha^\prime) {e^{\varphi} \over \sqrt{1 + e^{2 \varphi}
(C^{(0)})^2}} {\tilde {\cal F}} \right)|} \, \\
& & \\
& &+ \,\, {1 \over 6!}(2\pi\alpha^\prime) T_{{\rm B5}} \int d^6\xi
\,\,  \epsilon^{i_1 \dots i_6}
{\tilde {\cal G}}^{(6)}_{i_1 \dots i_6} \, .\\
\end{array}
\end{equation}
${\tilde {\cal F}}$ is 
defined as ${\tilde {\cal F}}=2\partial c^{(1)}+1/(2\pi\alpha^\prime)
C^{(2)}$ and 
${\tilde {\cal G}}^{(6)}$ 
is the gauge invariant Wess-Zumino term:
\begin{equation}
\begin{array}{rcl}
{\tilde {\cal G}}^{(6)} &=& \left\{
6\partial {\tilde c}^{(5)} - {1 \over 2\pi\alpha^\prime}{\tilde {\cal B}}
- {45 \over 2(2\pi\alpha^\prime)} 
{\cal B}C^{(2)}C^{(2)} -15 C^{(4)} {\tilde {\cal F}} 
\right. \\ & &\\
& & - 180 (2\pi\alpha^\prime) {\cal B} \partial c^{(1)}\partial c^{(1)}
-90 {\cal B} C^{(2)} \partial c^{(1)}
\\& &\\& &
\left.
+15 (2\pi\alpha^\prime)^2 {C^{(0)} \over e^{-2 \varphi} + (C^{(0)})^2} 
{\tilde {\cal F}}{\tilde {\cal F}}{\tilde {\cal F}}
\right\} \, .\\
\end{array}
\end{equation}
This action is S-dual to the D5-brane effective action, 
with the worldvolume fields
transforming as SL(2,Z) doublets. It is interesting to note that the
BI field transforms into a 1-form $c^{(1)}$, 
and not into a 3-form as one would
have expected. This could be useful for the construction of (p,q)
5-branes.

Analogously to this construction it is possible to derive the effective
action of the Type IIB KK-monopole by performing a T-duality
transformation in the effective action of the IIA NS-5-brane (see
\cite{EJL}). These results show explicitly the duality relations between
the various 5-branes of the IIA and IIB theories.

\vskip0.5cm
\noindent
{\large \bf Acknowledgements}

\smallskip
\noindent
Y.L. would like to thank the organizers of the 
32nd Symposium Ahrenshoop on the Theory of Elementary Particles,
Buckow (Germany), for giving her the opportunity to present this material.
The work of E.E. is part of the research program of the Dutch
Foundation FOM.


\end{document}